\documentclass[aip,reprint]{revtex4-1}
\usepackage{amsmath}
\usepackage{amsfonts}
\usepackage{amssymb}
\usepackage{natbib}
\usepackage{graphicx}
\usepackage{setspace}
\usepackage{color}
\usepackage{float}
\usepackage{xcolor}
\usepackage{ulem}
\def \s {\text{s}}
\def \ST {\text{ST}}
\def \sh {\text{sh}}

\def \cold {\text{cold}}
\def \f {\text{f}}
\def \h {\text{h}}
\def \p {\text{p}}
\def \n {\text{n}}
\def \s {\text{s}}
\def \b {\text{b}}
\def \h {\text{h}}

\def \sl {\text{sl}}

\def \rs#1 {{\bf #1}}
\def \tf#1 {#1}
\begin{document}

% Use the \preprint command to place your local institutional report number 
% on the title page in preprint mode.
% Multiple \preprint commands are allowed.
\preprint{}

\title{The Non Relativistic Interiors of Ultra-Relativistic Explosions:\\
Extension to the Blandford McKee Solutions} %Title of paper

% repeat the \author .. \affiliation  etc. as needed
% \email, \thanks, \homepage, \altaffiliation all apply to the current author.
% Explanatory text should go in the []'s, 
% actual e-mail address or url should go in the {}'s for \email and \homepage.
% Please use the appropriate macro for the type of information

% \affiliation command applies to all authors since the last \affiliation command. 
% The \affiliation command should follow the other information.

\author{Tamar Faran}
\email[]{tamar.faran@mail.huji.ac.il}
%\homepage[]{Your web page}
%\thanks{}
%\altaffiliation{}
\affiliation{Racah Institute of Physics, Hebrew University, Jerusalem 91904, Israel}

\author{Re'em Sari}
\affiliation{Racah Institute of Physics, Hebrew University, Jerusalem 91904, Israel}

% Collaboration name, if desired (requires use of superscriptaddress option in \documentclass). 
% \noaffiliation is required (may also be used with the \author command).
%\collaboration{}
%\noaffiliation

\date{\today}

\begin{abstract}
The hydrodynamics of an ultrarelativistic flow, enclosed by a strong shock wave, are described by the well known Blandford-McKee solutions in spherical geometry. These solutions, however, become inaccurate at a distance $\sim R/2$ behind the shock wave, where $R$ is the shock radius, as the flow approaches Newtonian velocities. In this work we find a new self-similar solution which is an extension to the Blandford-McKee solutions, and which describes the interior part of the blast wave, where the flow reaches mildly relativistic to Newtonian velocities. We find that the velocity profile of the internal part of the flow does not depend on the value of the shock Lorentz factor, $\Gamma$, and is accurate from $r=0$ down to a distance of $R/\Gamma^2$ behind the shock. Despite the fact that the shock wave is in causal contact with the entire flow behind it, a singular point appears in the equations. Nevertheless, the solution is not required to pass through the singular point: for ambient density that decreases slowly enough, $\rho \propto r^{-k}$ with $k<\frac{1}{2}(5-\sqrt{10})\cong0.92$, a secondary shock wave forms with an inflow towards the origin. 
\end{abstract}

\maketitle %\maketitle must follow title, authors, abstract and \pacs

% Body of paper goes here. Use proper sectioning commands. 
% References should be done using the \cite, \ref, and \label commands
\section{Introduction}
Self-similar solutions are obtained when a set of partial differential equations that contain both spatial and time derivatives can be reduced into a set of ordinary differential equations, characterized by a single similarity variable that is a combination of time and space variables. This similarity quality offers a great mathematical simplification and allows for a physical description of complex systems whose dynamics at late times are free of the spatial scales that existed in the initial conditions.
The most famous self-similar solution for the hydrodynamic equations, describing an explosion by the propagation of a strong shock wave is that of Taylor \cite{Taylor}, von Neumann \cite{Neumann} and Sedov \cite{Sedov}, known as the Sedov-Taylor solution. In their solution the shock wave propagates into a cold medium described by a power law density profile with a distance $r$ from the origin of the form $\rho \propto r^{-k}$. Assuming conservation of energy, they were able to find the scaling of the shock radius as a function of time. Such solutions, in which the shock is always in causal connection with the entire flow behind it are called first type solutions. Waxman \& Shvarts \cite{Waxman93} showed that if the density profile is steep enough ($k>3$), energy conservation considerations give the wrong scaling. Instead, the shock accelerates and loses causal connection with the flow, and a new scaling is found from the requirement that the solution passes through a singular point. These solutions are called second type solutions.

The ultra-relativistic analogue of the Sedov-Taylor problem was first treated by Blandford and McKee in 1976 \cite{BM76} (hereafter BM). They find first type self-similar solutions for a strong, spherical blast wave propagating into cold surroundings with a density profile of the form described above. The BM solution assumes that both the shock and the flow behind it retain ultra-relativistic Lorentz factors . However, the spatial velocity profile behind the shock drops significantly, and inevitably reaches trans-relativistic velocities. In this regime, the BM solution is no longer able to describe the dynamics of the flow.
While the BM solution probes only the relative vicinity of the shock, i.e., $\Delta r/R < 1$ where $R$ is the shock radius and $\Delta r = R-r$, our aim in this work is to find a self-similar solution that follows the entire flow, from the ultra relativistic regime ($\gamma\gg 1$) towards non-relativistic velocities ($\gamma\sim 1$). This solution, however, will not be able to accurately solve for the smallest scales behind the blast wave, where $\Delta r<R/\Gamma^2$, $\Gamma$ being the Lorentz factor of the shock.
We solve the hydrodynamic equations, without making any assumption about the Lorentz factor of the gas, and derive new self similar solutions for the dynamics of the flow also at large scales ($\Delta r> R$) behind the shock wave. In this era of high energy astrophysics, ultra-relativistic shock waves that give birth to the most energetic celestial events are of great interest. Most recently, the gravitational wave event, GW170817 was followed by an electromagnetic counterpart in all frequencies. Gamma rays from the event were detected by the \textit{Fermi} Gamma-ray Burst Monitor \citep{Goldstein17}, and the measured spectrum pointed to the existence of a relativistic shock breakout from the expanding ejecta. Late time radiation emerging from such an event is expected to be emitted from material moving at trans-relativistic and Newtonian velocities, which in some configurations can be described by the results of this work.

In section \ref{s:BM_solution} we review the properties of a strong ultra-relativistic shock wave as given in the solution of BM. We continue to derive the self-similar hydrodynamic  equations that describe the flow with the new similarity normalization in Section \ref{s:TR_continuation}, and solve them in Section \ref{s:TR_solution}. In Section \ref{s:causality} we discuss the causality of the shock with the flow behind it, and in Section \ref{s:BM_TR_connection} we show how the velocity in the TR solution connects to that of BM. The energy and the temperature of the shocked fluid in the self-similar solution are discussed in Sections \ref{s:energy}-\ref{s:temperature}. Finally, we discuss our results in Section \ref{s:discussion}.

\section{The Blandford-McKee Solution for a strong ultra-relativistic shock} \label{s:BM_solution}
The equations that describe the hydrodynamics of a relativistic fluid are represented by energy, momentum and particle number conservation:
\begin{subequations} \label{eq:hydro_ultra_rel}
    \begin{align}
       & \frac{\partial}{\partial t} \gamma^2(e+\beta^2p)+\frac{1}{r^\alpha} \frac{\partial}{\partial r} r^\alpha \gamma^2 \beta (e+p) = 0 \\
       &  \frac{\partial}{\partial t}  \gamma^2 \beta (e+p) + \frac{1}{r^\alpha} \frac{\partial}{\partial r} r^\alpha \gamma^2 \beta^2 (e+p) + \frac{\partial}{\partial r} p=0 \\
       & \frac{\partial}{\partial t} \gamma n + \frac{1}{r^\alpha} \frac{\partial}{\partial r} r^\alpha \gamma \beta n = 0 
    \end{align}
\end{subequations}
where $\alpha = 0,1,2$ refer to planar, cylindrical and spherical geometries, and $p$ and $e$ are the pressure and energy density in the fluid rest frame, respectively.  $\gamma$ is the Lorentz factor of the shocked fluid as seen in the upstream, and is related to the velocity $\beta$, by $\gamma = \frac{1}{\sqrt{1-\beta^2}}$. In this paper we will focus on the solutions in spherical geometry, i.e., $\alpha=2$.

BM solved the problem of an ultra-relativistic spherical blast wave following a strong shock, passing through an envelope of power-law density profile of the form
\begin{equation}
    \rho \propto r^{-k} ~,
\end{equation}
where the values of $k$ valid for first type solutions satisfy $k<17/4$ \citep{sari06}.

The jump conditions across the shock front follow from the conservation of energy ($w\gamma^2\beta$), momentum ($w\gamma^2\beta^2+p$) and particle number ($n\gamma \beta$) flux densities, where $w = e+p$ is the enthalpy in the fluid frame and $e$ is the energy density. Throughout this paper we assume an ultra-relativistic equation of state
\begin{equation}
    p = \frac{1}{3}e ~,
\end{equation}
corresponding to $\hat{\gamma}=4/3$, where $\hat{\gamma}$ is the ratio of specific heats. This assumption is valid as long as the internal energy is dominated by relativistic particles or radiation, and will break down once the particle rest mass can no longer be neglected relative to the pressure, i.e., $P\sim \rho$.

Under the assumption of an ultra-relativistic shock wave, BM found that the scale height of the shock is $R/\Gamma^2$, where $\Gamma$ is the Lorentz factor of the shock and $\Gamma\gg 1$. Since $R/\Gamma^2$ is the area of physical interest in their solution, they defined the following similarity variable
\begin{equation} \label{eq:chi}
    \chi = 1+2(m+1)\frac{R-r}{R/\Gamma^2} ~,
\end{equation}
where $m$ is given by: 
\begin{equation}\label{eq:gamma_shock}
    \frac{t\dot{\Gamma}}{\Gamma} = -\frac{m}{2} ~.
\end{equation}
Given this definition, the shock radius is
\begin{equation}
    R = t\bigg[1-\frac{1}{2(m+1)\Gamma^2}\bigg] ~.
\end{equation}
Using the similarity variable $\chi$ and the shock jump conditions, the self-similar functions are written as:
\begin{equation} \label{eq:gamma_def}
    \gamma^2(r,t) = \frac{1}{2}\Gamma^2(t)g(\chi)
\end{equation}
\begin{equation}\label{eq:p_def_UR}
    p(r,t) = P(t)f(\chi)
\end{equation}
and
\begin{equation}\label{eq:n_def_UR}
    n'(r,t) = N'(t)h(\chi)
\end{equation}
where $n'=\gamma n$ is the number density as measured in the upstream frame and we define
\begin{equation}\label{eq:BM_time_dependence}
    \frac{t \dot{P}}{P} = -m-k , ~\frac{t \dot{N'}}{N'} = -m-k ~.
\end{equation}
The boundary conditions
\begin{equation}\label{eq:BM_boundary_conditions}
    g(1)=f(1)=h(1)=1 ~
\end{equation}
ensure that the jump conditions at the shock front are satisfied. The condition for energy conservation, $E\sim n R^3 \Gamma^2$ yields the value of $m$,
\begin{equation} \label{eq:m}
    m = 3-k ~.
\end{equation}
Substituting equations \eqref{eq:gamma_def}--\eqref{eq:n_def_UR} with the boundary conditions in Eq \eqref{eq:BM_boundary_conditions} into equation set \eqref{eq:hydro_ultra_rel} gives the BM solutions for $g,f$ and $h$:
\begin{equation} \label{eq:BM_sol_g}
    g = \chi^{-1}
\end{equation}

\begin{equation}\label{eq:BM_sol_f}
    f = \chi ^{\frac{4k-17}{3(4-k)}} = \chi^{\alpha_f}
\end{equation}

\begin{equation}\label{eq:BM_sol_h}
    h = \chi ^{\frac{2k-7}{4-k}}  = \chi^{\alpha_h} ~,
\end{equation}
where we define $\alpha_f \equiv (4k-17)/[3(4-k)]$ and $\alpha_h \equiv (2k-7)/(4-k)$.
Since these are first type solutions, they are only valid for $k<17/4$.

Some intuition regarding the BM solutions can be gained by looking at Eq \eqref{eq:BM_sol_g}. At a distance $\Delta r$ behind the shock front the fluid moves at a velocity $\beta = 1-2(m+1) \frac{\Delta r}{R}$, where the constants account for the fact that the fluid is decelerating and are of order unity. It follows that the inner part of the solution cannot be highly relativistic. 
Even if the shock is ultra-relativistic at all times, far enough behind the shock front, the flow will transition to being mildly-relativistic due to the decreasing velocity profile. In the next section, we show that this also happens on a scale of $\chi \sim \Gamma^2$, corresponding to $\Delta r \sim R$. This regime was not treated in the work of BM, and is the focus of this paper.

\tf{For $4<k<17/4$, the BM solutions are hollow, meaning that the material is confined to the region $0<\chi<1$ and the flow never reaches Newtonian velocities. The solution found in this work therefore applies only to $k<4$.}

\section{Extension to the Blandford-McKee Solutions} \label{s:TR_continuation}
\subsection{The self-similar equations}
This work is focused on describing the flow at a distance $\Delta r\sim R$ behind the ultra-relativistic shock wave, where the BM solutions no longer hold, and the fluid transitions into the trans-relativistic regime.
Using equations \eqref{eq:gamma_def} and \eqref{eq:BM_sol_g}, we find that the coordinate $\chi_1$ where the flow reaches $\gamma=1$ is
\begin{equation}\label{eq:chi1}
    \chi_1 = \frac{\Gamma^2}{2} ~.
\end{equation}
We look for a similarity solution that describes the dynamics both in the ultra-relativistic and the non-relativistic regimes.
Since the trans-relativistic point with $\gamma\beta \sim 1$ is a special point in the solution, it must be at a fixed coordinate in the new self similar solution we are seeking. Hence, up to a constant, we normalize all the dependent variables by their value at this position.

We define the new self similar variable of the trans-relativistic solution:
\begin{equation}
    \xi = \frac{t}{r} ~,
\end{equation}
where the position of the shock is at $\xi\sim1$. We use this definition of the self-similar coordinate rather than the one defined in the Sedov-Taylor solution: $\xi_\ST = r/R$. We choose this inverse form to conform with the behaviour of $\chi$, such that both $\chi$ and $\xi$ increase behind the shock front.

\tf{The new solution we seek has a special velocity scale $c$, but describes also subrelativistic velocities. This is possible in a self similar solution with the similarity variable $\xi$, since velocity of a certain fraction of $c$ is obtained at a constant value of $\xi$. This is compatible with indications from the original BM solutions where the transition to mildly relativistic flow occurs roughly at $\chi\sim \Gamma^2$ or $r\sim R/2$.
}

While the BM solutions are accurate at $\chi_1$ only up to a constant factor, the time dependence of the hydrodynamic variables at $\chi_1$ is correctly given by the BM solution. Notice that the coordinate $\chi_1$ is itself time dependent in the original BM solution, but has a fixed $\xi$ in our new solution.
The time dependence of $f$ and $h$ at $\chi_1$ is
\begin{equation} \label{eq:f_chi1}
    f(\chi_1) = \bigg(\frac{\Gamma^2}{2}\bigg)^{\alpha_f} \propto t^{-m\cdot \alpha_f}
\end{equation}
\begin{equation} \label{eq:h_chi1}
    h(\chi_1) = \bigg(\frac{\Gamma^2}{2}\bigg)^{\alpha_h}\propto t^{-m\cdot \alpha_h} ~.
\end{equation}
The new hydrodynamic variables are therefore normalized to the BM solutions at $\chi_1$,
\begin{subequations}\label{eq:TR_var}
    \begin{align}
        &\beta = \bar{b}(\xi) \\
        & p = \Big[P(t)\cdot f(\chi_1)\Big]\bar{f}(\xi) = \bar{P}(t)\bar{f}(\xi)\\
        & n' = \Big[N'(t)\cdot h(\chi_1)\Big]\bar{h}(\xi)= \bar{N}'(t)\bar{h}(\xi)
    \end{align}
\end{subequations}
where we have from equations \eqref{eq:BM_time_dependence}, \eqref{eq:f_chi1} and \eqref{eq:h_chi1}
\begin{equation}
    \frac{t\dot{\bar{N}}'}{\bar{N}'} = \lambda_n, ~ \frac{t\dot{\bar{P}}}{\bar{P}} = \lambda_p ~
\end{equation}
and we defined
\begin{equation} \label{eq:lambda_p_n}
\begin{split}
&\lambda_p\equiv \frac{4k^2-20k+15}{3(4-k)}\\
&\lambda_n\equiv\frac{2k^2-10k+9}{4-k} ~.
% &\lambda_p\equiv -k-m(1+\alpha_f)\\
% &\lambda_n\equiv -k-m(1+\alpha_h) ~.
\end{split}
\end{equation}
In terms of the new variables, the time and spatial derivatives are:
\begin{equation}
        \frac{\partial}{\partial t} = \dot{\bar{P}}(t) \frac{\partial}{\partial \bar{P}}+ \dot{\bar{N}}'(t) \frac{\partial}{\partial \bar{N}'} + \frac{\xi}{t}(1-\beta \xi)\frac{\partial}{\partial \xi}
\end{equation}
and
\begin{equation}
    \frac{\partial}{\partial r} = -\frac{\xi^2}{t}\frac{\partial}{\partial \xi} ~.
\end{equation}
The boundary conditions for this solution are set at the shock, where $\chi=1$. \tf{In terms of $\chi$, $\xi$ can be written as
\begin{equation} \label{eq:chi2xi}
    \xi = 1+\frac{\chi}{2(m+1)\Gamma^2}~,
\end{equation}
% \begin{equation} \label{eq:chi2xi}
%     \chi = 1+2(m+1)\Gamma^2\Big(1-\frac{1}{\xi}\Big) ~.
% \end{equation}
so that to first order in $1/\Gamma^2$, the $\xi$ coordinate of the shock, corresponding to $\chi=1$, is
\begin{equation} \label{eq:xi_shock}
    \xi_\sh =  1 + \frac{1}{2(m+1)\Gamma^2}~.
\end{equation}
In the limit of $\Gamma\gg 1$, $\xi_\sh\approx 1$.
}

The new solution, with the similarity variable $\xi$ is unable to probe the immediate vicinity of the shock front. Instead, it describes the flow at a distance $\Delta r\gg R/\Gamma^2$ behind the shock. As will be shown in Section \ref{s:BM_TR_connection}, at this distance behind the shock the Lorentz factor of the flow does not depend on $\Gamma$, and the only information available about the shock is that $\Gamma\gg 1$. Accordingly, the only boundary condition that can be assigned at $\xi_\sh = 1$ is $\beta_\sh = 1$.

The functions $\bar{f}$ and $\bar{h}$ diverge at $\xi\rightarrow 1$. For that reason we define the following auxiliary functions,
\begin{equation} \label{eq:F_def}
    F(\xi) = \bar{f}(\xi)\bigg(1-\frac{1}{\xi}\bigg)^{-\alpha_f}
\end{equation}
and
\begin{equation} \label{eq:H_def}
    H(\xi) = \bar{h}(\xi)\bigg(1-\frac{1}{\xi}\bigg)^{-\alpha_h} ~,
\end{equation}
such that the following boundary conditions hold:
\begin{equation}\label{eq:boundary_b}
    \bar{b}(\xi\rightarrow 1) = 1 ~,
\end{equation}
\begin{equation}\label{eq:boundary_F}
    F(\xi\rightarrow 1) = \Big[4(m+1)\Big]^{\alpha_f} ~,
\end{equation}
\begin{equation}\label{eq:boundary_H}
    H(\xi\rightarrow 1) = \Big[4(m+1)\Big]^{\alpha_h} ~.
\end{equation}
We write the hydrodynamic equations in equation set \ref{eq:hydro_ultra_rel} in terms of $\beta$ and substitute the definitions in Eq \eqref{eq:TR_var}, yielding the new set of self-similar equations
\begin{equation}
\begin{split}
    &\frac{d \bar{f}}{d \xi}\Big((3+\bar{b}^2)-4\bar{b}\xi^2\xi\Big)+\bar{f}\Bigg[4\alpha \xi \bar{b}+(3+\bar{b}^2)\lambda_p+ \\
    &\frac{d \bar{b}}{d \xi}\bigg(2\bar{b}\xi\bigg(\frac{3+\bar{b}^2}{1-\bar{b}^2}+1\bigg)
    -4\xi^2-8\xi^2\frac{\bar{b}^2}{1-\bar{b}^2}\bigg)\Bigg]=0
    \end{split}
\end{equation}
\begin{equation}
    \begin{split}
         &\frac{d \bar{f}}{d \xi}\Big((3+\bar{b}^2)\xi-4\bar{b}\xi^2\Big)+\bar{f}\Bigg[\frac{d \bar{b}}{d \xi}\bigg(\frac{8 \bar{b} \xi}{1-\bar{b}^2}(\bar{b}-\xi) + 4\xi\bigg) \\
         & +4\lambda_p \bar{b}+4\alpha\bar{b}^2 \xi \Bigg] = 0
    \end{split}
\end{equation}
\begin{equation}
    \frac{d \bar{h}}{d \xi} \xi(1-\bar{b}\xi) + \bar{h}\bigg(\lambda_n+\alpha \bar{b} \xi - \frac{d \bar{b}}{d \xi}\xi^2\bigg)   = 0
\end{equation}
After rearranging the equations, we arrive at:
\begin{subequations} \label{eq:hydro_self_sim}
    \begin{align}
        &\frac{d \bar{b}}{d \xi} = -\frac{(\bar{b}^2-1)\big(\bar{b}^2(4\alpha+3\lambda_p)-4\alpha\bar{b}\xi -3\lambda_p\big)}{4\big(\bar{b}^2(3\xi^2-1)-\xi^2-4\bar{b}\xi+3\big)}  \label{eq:hydro_b}\\ 
        & \frac{d\bar{f}}{d \xi} = \frac{4 \alpha \bar{b} \xi (\bar{b} \xi -1) +\lambda_p(\bar{b}^2+2\bar{b}\xi-3)}{\xi(3-4\bar{b}\xi-\xi^2+\bar{b}^2(3\xi^2-1))}\bar{f} \equiv X_\f(\xi) \bar{f}  \label{eq:hydro_f}\\
        & \frac{d\bar{h}}{d\xi} = \frac{\bar{h}}{4(\bar{b}\xi -1)}\bigg(\frac{4 \lambda_n}{\xi}+\label{eq:hydro_h} \nonumber \\
        & \nonumber\frac{12 \alpha \bar{b}+3\lambda_p\xi +\bar{b}^4(4 \alpha+3\lambda_p)\xi -2\bar{b}^2(2 \alpha+3\lambda_p)\xi}{3-4\bar{b}\xi-\xi^2+\bar{b}^2(3\xi^2-1)} \\
        &+ \frac{4 \alpha\bar{b}^3(2\xi^2-1)}{3-4\bar{b}\xi-\xi^2+\bar{b}^2(3\xi^2-1)}\bigg) \equiv X_\h(\xi) \bar{h}  ~,
    \end{align}
\end{subequations}
where we defined $X_\f(\xi)$ and $X_\h(\xi)$ as the functions multiplying $\bar{f}$ and $\bar{h}$, respectively. Using equations \eqref{eq:F_def}, \eqref{eq:H_def}, \eqref{eq:hydro_f} and \eqref{eq:hydro_h}, we obtain the differential equations for $F$ and $H$:
\begin{subequations} \label{eq:hydro_FH}
\begin{align}
    &\frac{dF}{d\xi} = F(\xi)\bigg[X_f(\xi)-\alpha_f\frac{1}{\xi(\xi-1)}\bigg] ~,\\
    &\frac{dH}{d\xi} = H(\xi)\bigg[X_h(\xi)-\alpha_h\frac{1}{\xi(\xi-1)}\bigg] ~.
\end{align}
\end{subequations}

\subsection{The sonic line}
Self similar solutions of the second type are characterized by a singular point through which the solution is required to pass \citep{sari06}, while first type solutions do not contain such a singularity. Although the new solutions are an extension to the BM first type solutions, equation set \eqref{eq:hydro_self_sim} has a singularity point that satisfies
\begin{equation} \label{eq:xi_singular_line}
    \xi_{\text{singular}} = \frac{\sqrt{3}\pm\beta}{1\pm\sqrt{3}\beta} ~.
\end{equation}
This is the coordinate of a sound wave that satisfies $\xi = const$.
We can show this by writing the advective derivative of $\xi$,
\begin{equation} \label{eq:xi_dot}
    \dot{\xi} = \frac{\xi}{t}(1-\beta\xi) ~,
\end{equation}
which becomes $0$ when $\beta \xi = 1$.
The velocity of the $C_+$ and $C_-$ characteristics in the upstream frame is 
\begin{equation}
    \beta_\pm = \frac{d r_{\pm}}{dt} = \frac{\sqrt{3}\beta  \pm 1}{\sqrt{3} \pm \beta} ~,
\end{equation}
and the coordinate that satisfies $\beta_\pm \xi=1$ is
\begin{equation}
    \xi = \frac{\sqrt{3}\pm\beta}{1\pm\sqrt{3}\beta} ~,
\end{equation}
where the $+$ and $-$ signs correspond to the $C_+$ and $C_-$ characteristics, respectively. This result is equivalent to Eq \eqref{eq:xi_singular_line}. Within the range $1<\xi<\infty$, the condition $\beta_-\xi=1$ can be satisfied  only at $\xi=1, \beta=1$ (at the shock), and is therefore irrelevant.

The locus of points that satisfy $\beta_+\xi=1$ define the sonic line:
\begin{equation}
    \beta_\sl(\xi) = \frac{\xi-\sqrt{3}}{1-\sqrt{3}\xi} ~.
\end{equation}
In order for the solution to cross the sonic line, the numerator of Eq \eqref{eq:hydro_b} has to equal $0$. For the $C_+$ characteristic, this condition is satisfied both at $\xi=1, \beta=1$ and also at 
\begin{subequations} \label{eq:singuar_point}
    \begin{align}
        &\xi_\s = \frac{\sqrt{3}}{4}(4+\lambda_p) \\
        &\beta_\s = -\frac{\sqrt{3}\lambda_p}{8+3\lambda_p} ~.
    \end{align}
\end{subequations}
We denote this point as the singular point. For $k>3-\sqrt{3}/2\approx 2.13$, the singular point lies at $\xi<1$, outside the range of the solution. However, as we will show in Section \ref{s:temperature}, our new solution is inapplicable when $k>2$, in which cases the flow in the BM solution cools before becoming non-relativistic and the assumption of an ultrarelativistic equation of state breaks down. A major difference from second type solutions is that here the solution is not required to pass through the singular point. In Section \ref{s:secondary_shock_wave} we show that for certain values of $k$,  the solution does not pass smoothly through the singular point on the sonic line, $(\xi_\s,\beta_\s)$. Instead, a shock wave exists at some $\xi_{shock}$, where $1<\xi_{shock}<\xi_\s$, allowing the solution to pass from one side of the sonic line to the other without crossing it.
In the next section we obtain the solution for equation set \eqref{eq:hydro_self_sim} -- \eqref{eq:hydro_FH} for all relevant values of $k$.

\section{The solution of the hydrodynamic equations} \label{s:TR_solution}

\begin{figure}
\centering
\includegraphics[width=1\columnwidth]{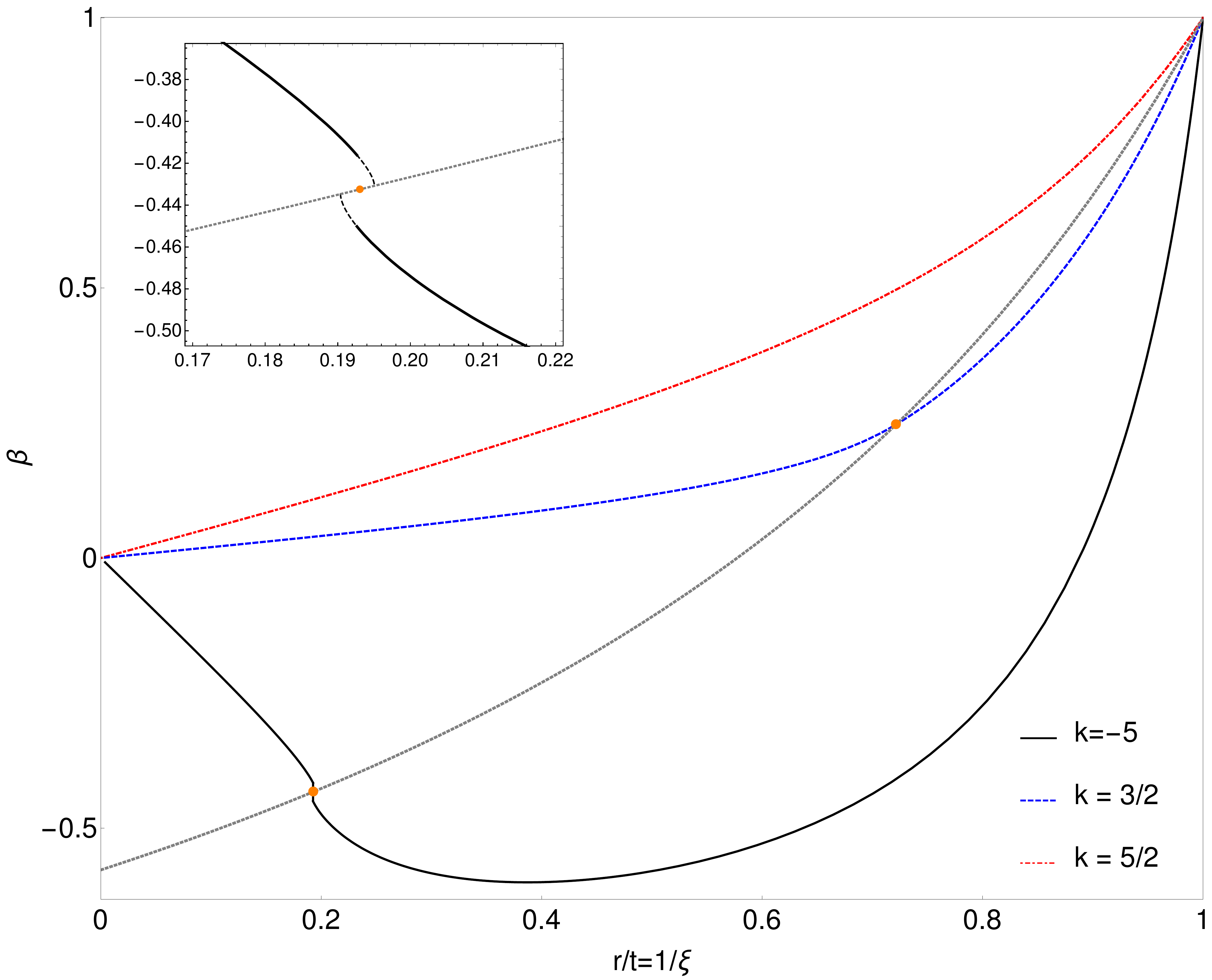}
\caption{The velocity $\beta$ for $k=-5$ (solid black), $k=1.5$ (dashed blue) and  $k=5/2$ (dot-dashed red). The inset focuses on the area around the singular point, shown in orange, in the $k=-5$ solution. This solution does not pass through the singular point, and at $r/R\sim 0.19$ a secondary shock is formed. The equations are solvable up to the singular line (dotted gray curve); beyond the secondary shock position the solution is plotted in dashed black. Despite their visual proximity, the position of the secondary shock does not coincide with the singular point. The two other curves demonstrate the case for which the solution crosses smoothly the sonic line ($k=3/2$) and the case in which the entire solution lies above the sonic line ($k=5/2$).
} \label{f:beta_k}
\end{figure}

\begin{figure}
\centering

\includegraphics[width=1\columnwidth]{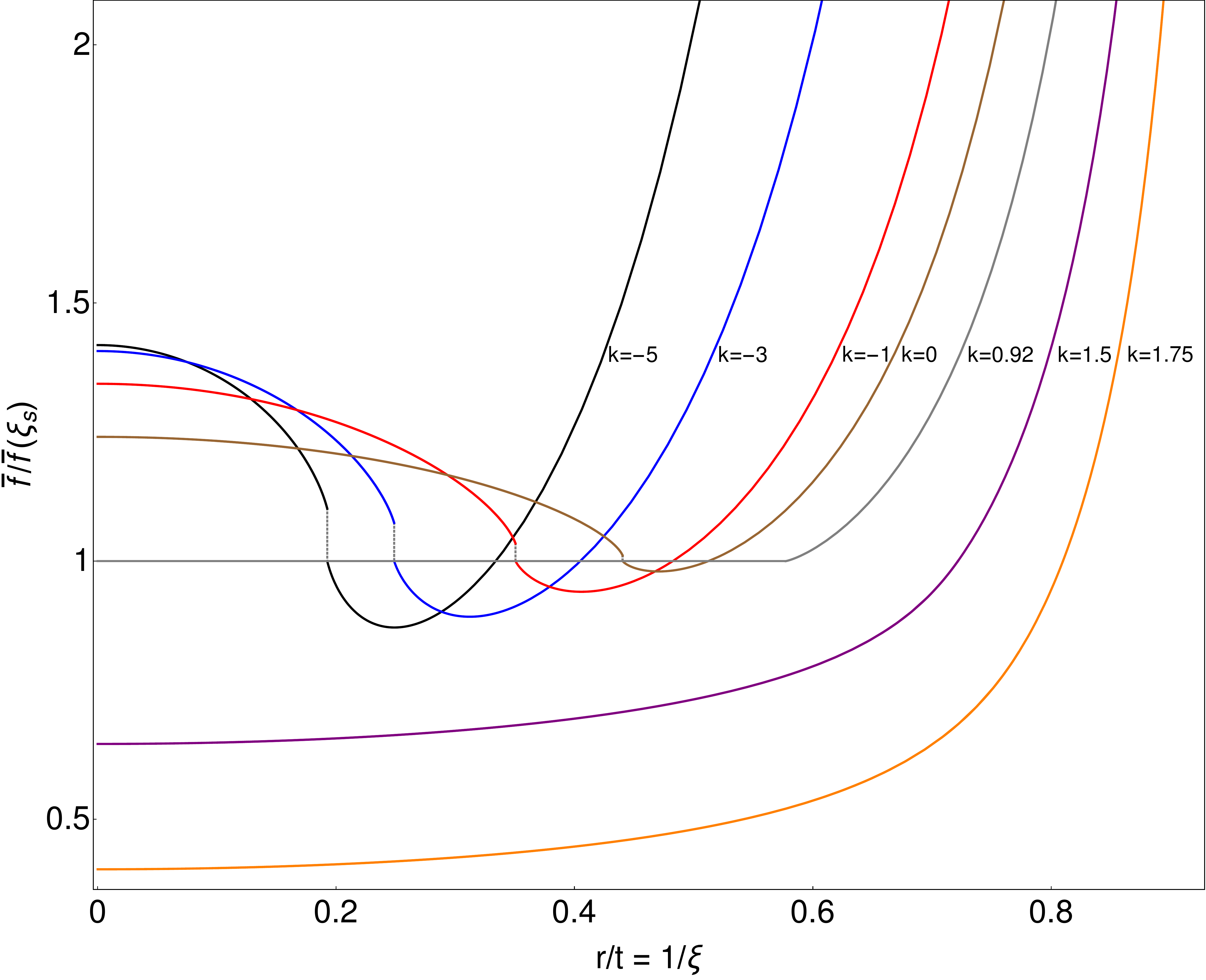} 
\caption{The profiles of $\bar{f}$ for various values of $k$ where $\Gamma=10^3$. \tf{The profiles are normalized to their values at the singular point. } The gray dotted lines indicate the secondary shock waves, which exist for $k<\frac{1}{2}(5-\sqrt{10})\cong 0.92$. At exactly $k=\frac{1}{2}(5-\sqrt{10})$, the pressure in the part of the solution internal to the singular point is identically constant.} \label{f:f_k}
\end{figure}

\begin{figure}
\centering
\includegraphics[width=1\columnwidth]{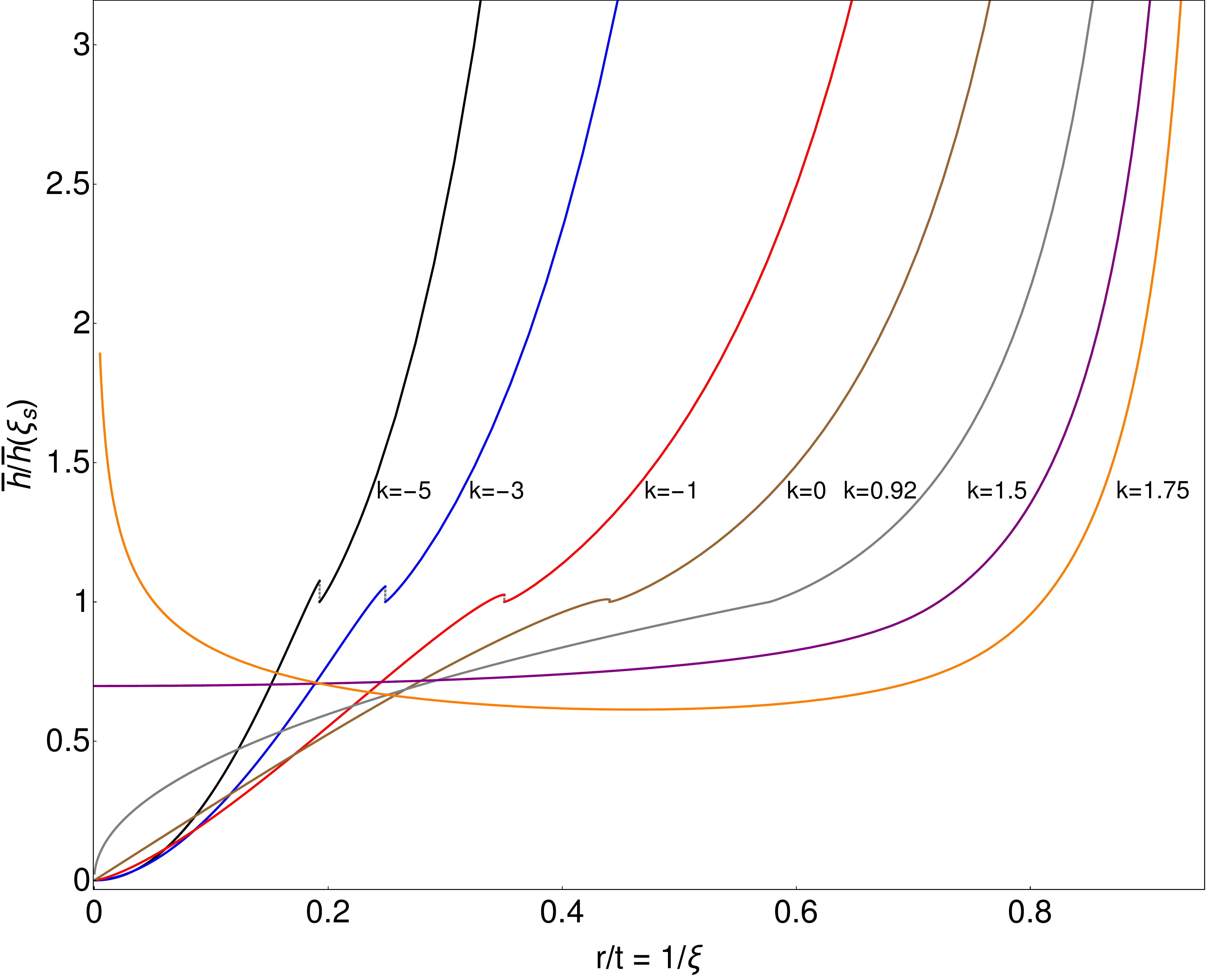} 
\caption{Same as Figure \ref{f:f_k} for $\bar{h}$ instead of $\bar{f}$.}\label{f:h_k}
\end{figure}

Equation sets \eqref{eq:hydro_self_sim} and \eqref{eq:hydro_FH} do not have an analytic solution, and must be solved numerically. Due to the existence of the singular point, we solve the equations in two directions. We start from the shock position towards $\xi_\s$, with the boundary conditions in equations \eqref{eq:boundary_b}--\eqref{eq:boundary_H}, and then solve from $r=0 ~(\xi\rightarrow\infty)$ towards $\xi_\s$. In Figures \ref{f:beta_k}--\ref{f:h_k} we show the solutions for $\bar{b}$, $\bar{f}$ and $\bar{h}$ for various values of $k$. A secondary shock exists when the solution does not pass smoothly through the singular point. The conditions for that are discussed in Section \ref{s:secondary_shock_wave}.
We plot the solutions for $\bar{b}$, $\bar{f}$ and $\bar{h}$ for various values of $k$ in Figures \ref{f:beta_k}--\ref{f:h_k}.

\subsection{The solutions in the limit $r\rightarrow0$} \label{s:solutions_at_r0}
\tf{At the origin we have $\beta(r=0) = 0$. } The behaviour of the velocity at $r \rightarrow 0$ ($\xi \rightarrow \infty$) is unique and depends only on the explicit time dependence of the pressure, $\lambda_p$. Taking the limit of Eq \eqref{eq:hydro_b} at $\xi\rightarrow \infty$, it simplifies to the differential equation
\begin{equation}
    \frac{d \bar{b}}{d \xi} \approx \frac{3\lambda_p+8\xi \bar{b}}{4 \xi^2} ~,
\end{equation}
whose solution is
\begin{equation} \label{eq:beta_center}
    \beta(\xi\gg 1) = -\frac{\lambda_p}{4}\xi^{-1} ~.
\end{equation}
This result can be understood as follows; the rate of change in the internal energy is a result of the work done on a fluid element per unit time, i.e.,
\begin{equation}
    \frac{d}{dt}\Big(\frac{4\pi}{3}r^3 3p\Big) \sim -4\pi r^2 p \beta ~.
\end{equation}
Substituting the definitions from Eq \eqref{eq:TR_var} and using the fact that $d\bar{f}/d\xi \rightarrow 0$ at the center (as we will show next), yields $\beta \sim -\lambda_p/4\xi^{-1}$. The normalization and sign of the velocity at the center are determined only by the time dependence of the pressure at fixed $\xi$.
The relation between the sign of the velocity and $\lambda_\p$ can be understood intuitively. When $\beta>0$, the fluid expands, and the pressure decreases. Since the pressure profile is constant at $r\rightarrow0$, $\lambda_\p$ must be negative. When $\beta<0$, the fluid is compressed and the pressure increases, which requires $\lambda_\p>0$. According to this result, negative velocities near the origin, corresponding to $\lambda_\p>0$ are achieved for $k<\frac{1}{2}(5-\sqrt{10})\cong 0.92$.

In the BM solution, $k=0$ and therefore $\lambda_p>0$. However, it is well known that the pressure behind an ultra-relativistic shock wave propagating into a uniform medium decreases with time. Since the shock decelerates, a constant value of $\chi$ refers to an increasing ratio of $r/R$, and the pressure increases accordingly. The result of Eq \eqref{eq:beta_center} is general, since equation set \eqref{eq:hydro_self_sim} is applicable in all regimes of $\gamma$. For example, in the Sedov-Taylor solutions,
\begin{equation}\label{eq:beta_ST}
    u^\ST(r\rightarrow0) = \alpha_\ST\frac{r}{t}\frac{1}{\hat{\gamma}} = -\frac{1}{4}\frac{d \log p}{d \log t}\frac{r}{t} ~,
\end{equation}
where $u^{\ST}$ is the velocity in the Sedov-Taylor solution and $\alpha_\ST \equiv 2/(5-k)$ is defined by $R\propto t^{\alpha_\ST}$. In the second equality we substituted $\hat{\gamma} = 4/3$. Since $d\log p/d\log t$ at constant $\xi$ can be interpreted as $\lambda_\p$, equations \eqref{eq:beta_center} and \eqref{eq:beta_ST} are equivalent.  Note, however, that while in our new solution $\beta$ can have either positive or negative values near the origin, the velocity in the Sedov-Taylor solution is always positive around the origin.

The pressure at the center approaches a constant value
\begin{equation}
    \bar{f}(\xi\rightarrow\infty) = \bar{f}_0~,
\end{equation}
where $\bar{f}_0$ is an unknown constant, and the density at the center is a power-law and satisfies
\begin{equation}\label{eq:h_center}
    \bar{h}(\xi\gg 1)=\bar{h}_0 \bigg(\frac{\xi}{\xi_0}\bigg)^{\frac{3\lambda_p-4\lambda_n}{4+\lambda_p}} ~.
\end{equation}
Using the definitions of $\lambda_\p$ and $\lambda_\n$ in Eq.\eqref{eq:lambda_p_n}, we find that the density near the origin increases for $k>3/2$ and decreases to $0$ otherwise. A constant density profile forms near the center for $k=3/2$, as shown in Figure \ref{f:h_k}.

\subsection{The secondary shock wave} \label{s:secondary_shock_wave}
For certain values of $k$ the solution does not pass the sonic line smoothly through the singular point. Instead, a secondary outgoing shock wave is formed, which allows the solution to pass from one side of the sonic line to the other. To quantify the condition on the value of $k$ where a transition from a continuous solution to a secondary shock occurs, we analyze the vicinity of the singular point. We expand Eq \eqref{eq:hydro_b} to first order in $\xi$ around $\xi_s$, and assume that the solution passes through the singular point. This implies that
\begin{equation}
\begin{split}
    \frac{d \beta}{d \xi}\bigg|_{\xi_s} = \frac{\beta_s}{\xi_s} \approx \frac{64 \lambda_p +24\lambda_p^2\pm 4\sqrt{-2\lambda_p(8+\lambda_p)(8+3\lambda_p)^2}}{(8+3\lambda_p)^3}~,
    \end{split}
\end{equation}
where the first equality applies the assumption that the solution goes through $\xi_\s$.
The above expression is real only when $\lambda_p<0$, which translates to $k>\frac{1}{2}(5-\sqrt{10}) \cong 0.92$. For these values of $k$, the solution crosses the sonic line smoothly through the singular point and no shock wave is formed. As we showed in Section \ref{s:solutions_at_r0}, this is also the condition for positive velocities around $r\rightarrow0$. The formation of the secondary shock thus coincides with inflow near the origin.

The secondary shock wave is a weak shock due to the high pressure that exists in the downstream after the passage of the initial blast wave. Moreover, the fluid has a non-zero velocity, which makes the jump conditions across the secondary shock different from those of the original shock.

Let us denote the hydrodynamic values in the immediate upstream of the secondary shock by the subscript $_1$, and the ones in the immediate downstream by the subscript $_2$. Given these notations, the jump condition on the Lorentz factor is
\begin{equation}\label{eq:gamma_jump}
\gamma_{2,\sh} = 3\sqrt{\frac{\gamma_{1,\sh}^2-1}{8\gamma_{1,\sh}^2-9}} ~,
\end{equation}
and the conditions on the pressure and density, respectively are
\begin{equation}
    p_2 = \frac{1}{3}\Big(8\gamma_{1,\sh}^2-9\Big)p_1
\end{equation}
and
\begin{equation}
    \rho_2 = \frac{\sqrt{8\gamma_{1,\sh}^4-17\gamma_{1,\sh}^2+9}}{\gamma_{1,\sh}} \rho_1 ~,
\end{equation}
where $\gamma_{1,\sh}$ and $\gamma_{2,\sh}$ are the upstream and downstream Lorentz factors of the fluid in the shock reference frame, respectively.

We cannot know a priori the conditions of the pressure and density at $r\rightarrow 0$ ($\xi\rightarrow\infty$). \tf{Nevertheless, as explained in Section \ref{s:solutions_at_r0}, the velocity near the origin must approach $0$. }
It is thus possible to solve Eq \eqref{eq:hydro_b} from $r=0$ up to the point where the solution encounters the singular line. The position of the secondary shock wave, $\xi_{\sh,\text{sec}}$, is found by solving Eq \eqref{eq:gamma_jump} for $\xi$, and will always satisfy $\xi_{\sh,\text{sec}}<\xi_s$ since the shock has to be supersonic. Once the value of $\gamma_{2,\sh}$ is known, $p_2$ and $\rho_2$ can also be found and equations \eqref{eq:hydro_f} and \eqref{eq:hydro_h} are solved from $\xi_{\sh,\text{sec}}$ towards $\xi\rightarrow\infty$.

\subsection{Analytic solution for $k=\frac{1}{2}(5-\sqrt{10})$}

For $k=\frac{1}{2}(5-\sqrt{10})\cong 0.92$, it is easy to verify by substitution that the solution $p=constant$ and $\beta = 0$ satisfies equation set \eqref{eq:hydro_self_sim}.
This solution also satisfies the boundary condition at $\xi=\infty$, but not the boundary condition at $\xi=1$. We therefore argue that it applies for $\xi_\s<\xi$.
For this value of $k$,  we also have $\lambda_\p=0$ such that the pressure is constant in time. As discussed in Section \ref{s:secondary_shock_wave}, this is the same value of $k$ below which a secondary shock wave forms. Since this is also the value of $k$ separating between inflow and outflow around the origin, it is not surprising that for $k=\frac{1}{2}(5-\sqrt{10})$ the velocity is exactly $0$ in the part internal to the singular point. By substituting $\lambda_\p=0$ and $\bar{b}=0$ in Eq \eqref{eq:hydro_h}, it is easy to see that the density at $\xi_\s<\xi$ is a power law and is equal to
\begin{equation} \label{eq:h_center_k092}
    \bar{h} = \bar{h}_\s \bigg(\frac{\xi}{\xi_\s}\bigg)^{-\lambda_\n} ~,
\end{equation}
where $\bar{h}_\s$ is the value of $\bar{h}$ at $\xi_\s$, and $\lambda_\n =3(-3+\sqrt{10}) \cong 0.49$ for this value of $k$. That thee density should be a power law with a power of $-\lambda_\n$ is not surprising. Since the pressure profile is constant in space and time, a fluid element, which stays at a fixed $r$ position due to the vanishing velocity, cannot be compressed. The only solution that would satisfy this is the one in Eq \eqref{eq:h_center_k092}.
We show the solution for $\bar{f}$ and $\bar{h}$ in this case in Figures \ref{f:f_k} and \ref{f:h_k}, respectively.

\section{Causality with the shock}\label{s:causality}

Here we examine the causality of the shock with respect to the flow behind it, by inspecting the solution in the $(r/t,\beta)$ plane. 
The  velocity at the origin vanishes, $\beta=0$, and thus the origin is represented by the $(0,0)$ point. The sonic line at $r=0$, satisfies $\beta_\sl=-\sqrt{3}/3$.
Therefore the interior of the solution resides above the sonic line and is subsonic.
At the other end, $\xi=1$, our boundary condition implies $\beta=\beta_{sl}=1$, and the solution starts on the sonic line.

If $\beta$ decreases inwards faster than $\beta_\sl$, the solution will have to pass back to the other side of the singular line.
Taking the derivative of $\beta_\sl$ at $\xi=1$, we have
\begin{equation}
\frac{d \beta_\sl}{d\xi}\bigg|_{\xi = 1} = -\frac{2}{(1-\sqrt{3})^2} ~,
\end{equation}
while
\begin{equation}
    \frac{d \beta}{d\xi}\bigg|_{\xi = 1} = \frac{1}{4}\Big(-12-3\lambda_\p-\sqrt{160+72\lambda_\p+9\lambda_\p^2}\Big) ~.
\end{equation}
 For $3-\sqrt{3}/2<k<4$ the flow throughout the solution lies above the sonic line. However, for $k<3-\sqrt{3}/2\cong 2.13$, $\beta$ goes below the singular line and will therefore have to either cross it at the singular point, which occurs if $\frac{1}{2}(5-\sqrt{10})<k<3-\sqrt{3}/2$, or jump across it through a shock, if $k<\frac{1}{2}(5-\sqrt{10})$. These three possibilities are shown in Figure \ref{f:beta_k}.
Eq. \eqref{eq:xi_dot} implies that the flow behind the shock is divided by the sonic line to a region where $C_+$ characteristics move away from the shock ($\beta_+\xi<1$) and towards the shock ($\beta_+\xi>1)$. Thus, for $k<3-\sqrt{3}/2$ all $C_+$ characteristics approach $\xi_\s$ asymptotically with time, never reaching $\xi=1$.
Naively, this result is contradictory to the BM solution, where the shock is in causal connection with the flow behind it, and the coordinate of the $C_+$ characteristic, $\chi_+$, satisfies $d\chi_+/dt<0$ everywhere between the shock and the sonic line (see equation 36 in \citealt{best_sari}), and reaches the shock at $\chi = 1$ in finite time. This may seem as a paradox, since our solution agrees with the BM solution for $2\lesssim\chi\lesssim\Gamma^2$, and therefore the fluid within a certain distance behind the shock must be in causal connection with it also in terms of the new $C_+$ coordinate, $\xi_+$.
We can investigate this contradiction by writing $d\chi_+/dt$ in terms of $\xi_+$ and $\beta_+$:
\begin{equation}
    \frac{d \chi_+}{dt} = 2(m+1)\frac{\Gamma^2}{t} \bigg[\frac{1-\beta_+\xi_+}{\xi_+} - m\Big(1-\frac{1}{\xi_+}\Big)\bigg] ~.
\end{equation}
The first term inside the square brackets is equivalent to Eq \eqref{eq:xi_dot} and has a positive contribution to $d\chi_+/dt$ for $k<3-\sqrt{3}/2\cong 2.13$ and $\xi<\xi_\s$. However, the second term comes from the derivative of $\Gamma$ with time, and when the shock decelerates ($m>0 \leftrightarrow k<3$) has a negative contribution which is larger than the first term.
It is therefore misleading to think of the shock front as if it is located at $\xi=1$. The shock is located at $\xi \sim 1+1/\Gamma(t)^2$, which increases with time for $m>0$. The contradiction is therefore settled; $C_+$ characteristics move away from $\xi=1$, but the shock catches up with them due to its deceleration.

In terms of the BM solution, $d\chi_+/dt$ can be written as:
\begin{equation}
    \frac{d \ln \chi_+}{d \ln t} = (m+1)\Big[2(2-\sqrt{3})-1\Big] ~.
\end{equation}
This equation, together with the time dependence of $\Gamma$ (Eq \ref{eq:BM_time_dependence}) can be solved to find the maximal coordinate, $\chi_0$, from which a sounds wave arrives at the shock wave before $\Gamma=1$ according to Eq \eqref{eq:BM_time_dependence}:
\begin{equation} \label{eq:dchidt_BM}
    \chi_0 = \Gamma_0^{-\frac{2(m+1)(3-2\sqrt{3})}{m}}~.
\end{equation}
The expression in the power of $\Gamma$ is smaller than $2$ for $k<3-\sqrt{3}/2\cong 2.13$. Therefore, for $k<3-\sqrt{3}/2$, $\chi_0<\Gamma^2$ and information from the Newtonian interior does not arrive at the shock before it becomes non-relativistic.

\begin{figure}
\centering
\includegraphics[width=1\columnwidth]{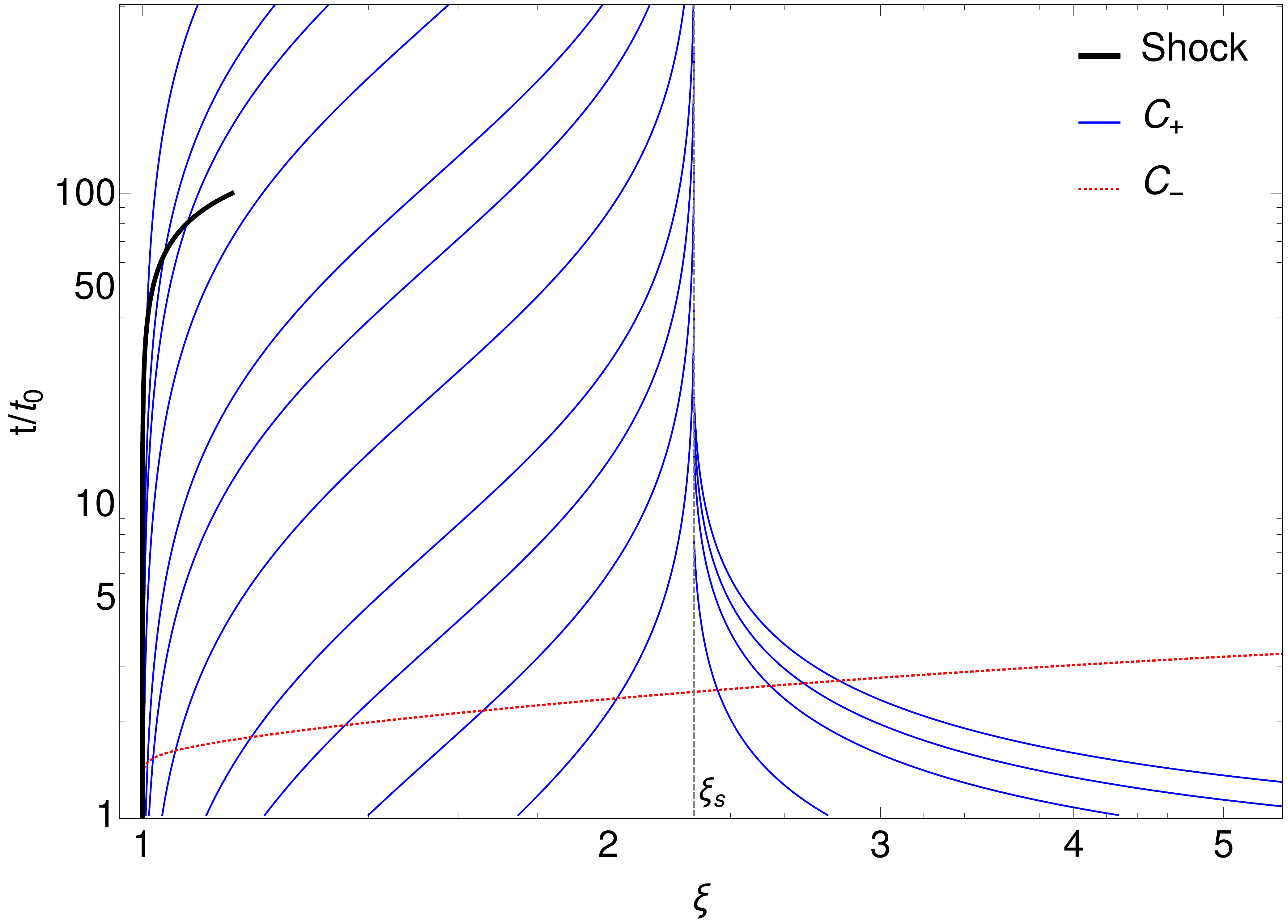}
\caption{Thin blue lines: $C_+$ characteristics emerging from different positions behind the shock wave in the case of $k=0$. The initial Lorentz factor of the shock wave (thick black line) at $t=t_0$ is $\Gamma_0 = 10^3$, and it initially lies at $\xi\sim 1$. We plot the shock position up to the time when $\Gamma=1$ according to Eq \eqref{eq:gamma_shock}. All $C_+$ characteristics approach the singular point (dashed-gray vertical line) with time, and never reach $\xi=1$. Nevertheless, the shock broadens due to its deceleration, and some $C_+$ characteristics emerging from $\xi<\xi_\s$ cross the shock position before it becomes non-relativistic, despite their increasing $\xi$ coordinate. \tf{We also show the $C_-$ characteristic emerging from the shock position $\xi \sim 1+1/\Gamma_0^2$ at $t_0$. A sound wave moving away from the shock reaches the origin at $t/t_0 \simeq 2.7 ~ \Gamma_0^{0.07}$, before the shock becomes non-relativistic.}
} \label{f:Cplus_char_shock}
\end{figure}

In Figure \ref{f:Cplus_char_shock} we show the $C_+$ characteristics emerging from different positions behind the shock, for an initial Lorentz factor of $\Gamma_0 = 10^3$ and $k=0$. $C_+$ characteristics emerging from $\xi<\xi_\s$ propagate backwards, while $C_+$ characteristics emerging from $\xi>\xi_\s$ propagate forward, all moving asymptotically towards $\xi_\s$. While none of the characteristics ever reaches $\xi=1$, the shock decelerates and crosses some of them after finite time. The shock reaches $\xi \sim \frac{4(m+1)}{4(m+1)-1}$ upon becoming non-relativistic, independent of the initial $\Gamma$. Beyond this point in time, the new self-similar solution is no longer correct. \tf{We also show one $C_-$ characteristic emerging from the shock position, which propagates all the way to the origin in finite time.}

\section{Connecting to the BM solution} \label{s:BM_TR_connection}
A surprising feature of the BM solutions is that the velocity does not depend on the value of $\Gamma$ at a given distance $\Delta r\gg R/\Gamma^2$ from the shock. This can be easily understood if we write the ultra-relativistic solution for $\gamma^2$ in this limit,
\begin{equation}
\begin{split}
    \gamma^2(\Delta r) = &\frac{1}{2}\Gamma^2\big[\chi(\Delta r)\big]^{-1} = \frac{1}{2}\Gamma^2\bigg[1+2(m+1)\frac{\Delta r}{R/\Gamma^2}\bigg]^{-1} \approx \\
    & \approx\frac{1}{4(m+1)}\frac{R}{\Delta r} ~,
    \end{split}
\end{equation}
which is independent of $\Gamma$ in the limit $\Gamma\gg 1$. 
This result can be understood intuitively when writing the expression for the distance a fluid element obtains from the shock after being shocked at $t_\s$:
\begin{equation}
    \Delta r = \int_{t_\s}^t c(1-\beta)dt \propto \frac{c t}{2\gamma^2}~,
\end{equation}
where the proportionality assumes that $t \gg t_\s$ and $\gamma\gg 1$,  while the proportionality constant depends on how $\gamma$ evolves with $t$. Since the Lorentz factor of a fluid element decreases with time, most of the deceleration occurs between $t_\s$ and $2t_\s$, and the fluid reaches the same velocity regardless of $\Gamma(t_\s)$.
This result implies that the boundary condition of $\beta=1$ ($\gamma\rightarrow \infty$) is sufficient to make the velocity in the new solution consistent with that of BM. The only manifestation of the shock Lorentz factor in this context is the location where the new solution departs from the BM solution. 
Nevertheless, the velocity in the Newtonian extension is not entirely ignorant about the shock wave, since Eq \eqref{eq:hydro_b} contains the value of $m$ which accounts for its deceleration rate.

The BM solutions describe the region $\Delta r \sim R/\Gamma^2$ behind the shock. This distance corresponds to $\chi\sim 1+2(m+1)$ according to Eq \eqref{eq:chi}. Using the relation between $\chi$ and $\xi$ (Eq \ref{eq:chi2xi}), the Newtonian extension can only be accurate at $\xi$ values larger than $\xi\sim 1+1/\Gamma^2$. At values of $\xi$ smaller than that, the dynamics depend on $\Gamma$ and are correctly described by the BM solution.

\begin{figure}
\centering
\includegraphics[width=1\columnwidth]{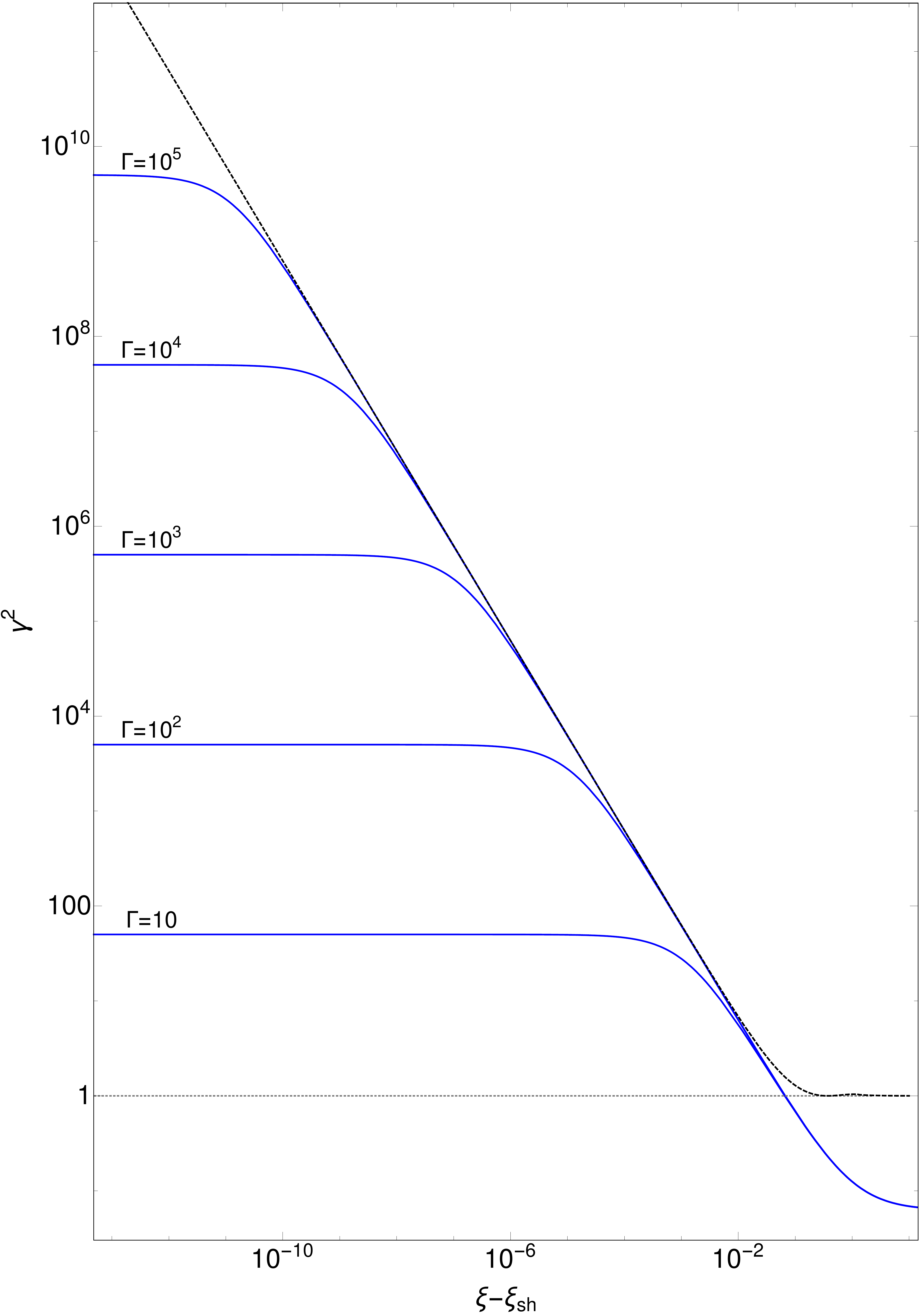} 
\caption{The solution for $k=0$ (dashed black line) which is independent of the value of $\Gamma$, and the BM solutions for different values of $\Gamma$ (solid blue lines). \tf{The x-axis is the distance from $\xi_\sh$, as defined in Eq \eqref{eq:xi_shock}. } The BM and the Newtonian extension solutions depart at $\xi-1 < 1/\Gamma^2$, where the latter fails to describe the ultra-relativistic dynamics in the vicinity of the shock.
From the other end, the BM solutions fail when $\gamma\sim 1$, which occurs for $k=0$ at $\xi\sim 1.07$. Internal to this point, only the new solution is valid.} \label{f:gamma2_near_shock}
\end{figure}

In Figure \ref{f:gamma2_near_shock} we show $\gamma^2$ as a function of $\xi-\xi_\sh$ in the Newtonian extension solution for $k=0$, which is independent of $\Gamma$, together with the BM solutions for various values of $\Gamma$. Close to the shock ($\xi-\xi_\sh\lesssim 1/\Gamma^2$), the Newtonian extension clearly does not trace the true velocity of the fluid, and approaches infinite Lorentz factors towards $\xi-\xi_\sh\rightarrow0$.  The BM solution becomes inaccurate as $\gamma$ tends to $1$. This happens around $\chi_1 = \Gamma^2/2$ where the corresponding $\xi$ coordinate depends also on the value of $k$. For $k=0$, this occurs at $\xi-1\sim 0.07$, as seen in Figure \ref{f:gamma2_near_shock}. The BM solutions are therefore accurate only over a small fraction of the volume.

\section{The energy in the solution} \label{s:energy}
Although the new solution found in this work is a continuation of a first type ultra-relativistic solution, energy is not conserved in the region $1<\xi<\infty$. This can be seen by writing the expression for the integrated energy,
\begin{equation} \label{eq:energy_itegral}
\begin{split}
    E &= \int_{r_1}^{r_2} \gamma^2(e+p)4 \pi r^2 dr = \\
    &= 16\pi \bar{P}(t) t^{3} \int_{\xi_1}^{\xi_2} \frac{\bar{f}(\xi)}{1-\bar{b}(\xi)^2} \xi^{-3}d\xi \\
    & \propto  I(\xi_1, \xi_2)\cdot t^{3+\lambda_p} ~,
    \end{split}
\end{equation}
where $I(\xi_1,\xi_2)$ represents the result of the integration over $\xi$. Generally, $3+\lambda_p \neq 0$, such that the energy of the self-similar solution between any two fixed values of $\xi$ can increase or decrease with time. In addition, the energy,  written in this form, diverges at $\xi\rightarrow 1$, where $\beta=1$. These two `artifacts' stem from the fact that our solution does not describe the shock itself, and are not a manifestation of the inner, non-relativistic part of the solution. We will now show that by applying a time-dependent lower boundary condition, which represents the true position at the shock, this divergence is eliminated and the energy is constant.

As explained in Section \ref{s:causality}, the shock position does not remain at a constant $\xi$ in time, but is located at $\chi= 1$, or $\xi \sim 1+1/\Gamma^2$. Since most of the energy is concentrated close to the shock, it can then be estimated as

\begin{equation} \label{eq:energy_approximation}
    E \approx \Gamma^2 P \Delta V = 4\pi R^2 \Gamma^2 P \Delta r \propto t^{-m} \Delta \xi ~,
\end{equation}
where $\Delta V=4\pi R^2 \Delta r$ and $\Delta \xi$ is the scale over which the energy changes significantly, corresponding to $\Delta\chi\sim 1$:
\begin{equation}
    \Delta \xi = \frac{1}{\Gamma^2}\propto t^m ~.
\end{equation}
Plugging $\Delta \xi$ back into Eq \eqref{eq:energy_approximation}, we find that the energy is indeed constant and finite when integrated over the correct region.

For the sake of discussion, it is worthwhile looking at the behaviour of the energy at some fixed range of $\xi$ as indicated by Eq \eqref{eq:energy_itegral}. The energy in the new self-similar solution increases with time for $k<3$, and decreases for $k>3$. In order to understand this, we should remember that the self-similar variables are normalized to the BM solution at $\chi_1 = \Gamma^2/2$. In the former case, $m>0$ and the shock decelerates. $\chi_1$ now represents a shorter distance behind the shock, causing the energy at this point to increase with time. Hence, the value of $p$ at constant $\xi$ increases. Accordingly, for $k>3$ the shock accelerates and less energy is confined in the internal region than is in the ultra-relativistic part such that the energy at constant $\xi$ decreases.
It is interesting to point out that the energy is constant in time for $k=3$, since in that case the shock does not collect mass, and keeps a constant Lorentz factor.

In the relativistic second type solutions \citep{sari06, best_sari}, the total energy cannot increase, reflecting the fact that the flow in the outer region is independent of that in the inner region.
Since most of the energy in the second type solutions is located behind the singular point ($\chi >\chi_\s$), no energy can transfer from the inner to the outer region. In our new solution, most of the energy, formally infinity, resides near the shock, and characteristics moving backwards allow for the increase of energy.
This allows the energy in a fixed range of $\xi$ to increase, which is impossible in second type solutions.

\section{The temperature behind the shock} \label{s:temperature}
Both in the Newtonian extension and the BM solutions an ultra-relativistic equation of state is applied, assuming that the shocked fluid is hot, i.e., $p\gg\rho$ at all times. Generally, the expanding fluid can cool through adiabatic expansion and when $P/\rho < 1$ the fluid temperature becomes non-relativistic. In the BM solutions, the temperature profile behind the shock always decreases for $k<4$. Since the new solution assumes that when the gas reaches $\gamma\sim 1$ it is still hot, we need to consider only cases in which the self-similar coordinate where the gas becomes cold, $\chi_\cold$ satisfies $\chi_\cold>\chi_1$. 

Using the definitions in Eq \eqref{eq:gamma_def}--\eqref{eq:n_def_UR} and the BM solutions in Eq \eqref{eq:BM_sol_g}--\eqref{eq:BM_sol_h} we find
\begin{equation}
    \chi_\cold \sim \Gamma ^{\frac{6(4-k)}{k+4}} ~.
\end{equation}
Recalling that $k<4$, the power law index is always positive, such that $\chi_\cold$ decreases with time and the fluid cools towards the shock, reflecting the fact that fluid elements at larger distances from the shock have had more time to cool.
Using Eq \eqref{eq:chi1}, we can write the ratio between $\chi_1$ and $\chi_\cold$:
\begin{equation}
    \frac{\chi_1}{\chi_\cold}\sim \Gamma^{\frac{8k-16}{k+4}} ~,
\end{equation}
showing that the fluid slows down before cooling only if $k<2$. In the rest of the cases, our solution may not be valid, as the equation of state may be irrelevant.

In the classical case of the BM solution for $k=0$,
\begin{equation}
    \chi_\cold\big |_{k=0} \sim \Gamma^6 ~.
\end{equation}
Using Eq \eqref{eq:chi1}, the ratio between $\chi_1$ and $\chi_\cold$ is
\begin{equation}
    \frac{\chi_1}{\chi_\cold} \Bigg |_{k=0} \sim \Gamma^{-4} ~,
\end{equation}
Therefore, as long as the shock has not slowed down significantly and $\Gamma\gg1$, fluid elements slow down before they cool, and we can assume that the fluid is hot at least down to where $\gamma=1$.

In order to test whether the temperature profile of the shocked fluid increases or decreases around the origin, we write the temperature in the internal solution:
\begin{equation}
    T\sim \frac{P}{\rho} = \frac{\bar{P}(t)}{\bar{N}'(t)}\frac{\bar{f}(\xi)}{\bar{h}(\xi)/\gamma} = \frac{\bar{P}(t)}{\bar{N}'(t)}\frac{\bar{f}(\xi)}{\bar{h}(\xi)}\frac{1}{\sqrt{1-\bar{b}^2(\xi)}} ~.
\end{equation}
Far behind the shock, $\xi\rightarrow \infty$, $\bar{b}\rightarrow 0$ and $\bar{f}\rightarrow \bar{f}_0$:
\begin{equation}
    T\sim  \frac{\bar{P}(t)}{\bar{N}'(t)}\frac{\bar{f}_0}{\bar{h}(\xi)}~.
\end{equation}
The temperature profile is inversely proportional to the density towards $r\rightarrow0$, thus it increases for decreasing density profiles, a condition  satisfied for $k<3/2$, and decreases otherwise.

\begin{figure}
\centering
\includegraphics[width=1\columnwidth]{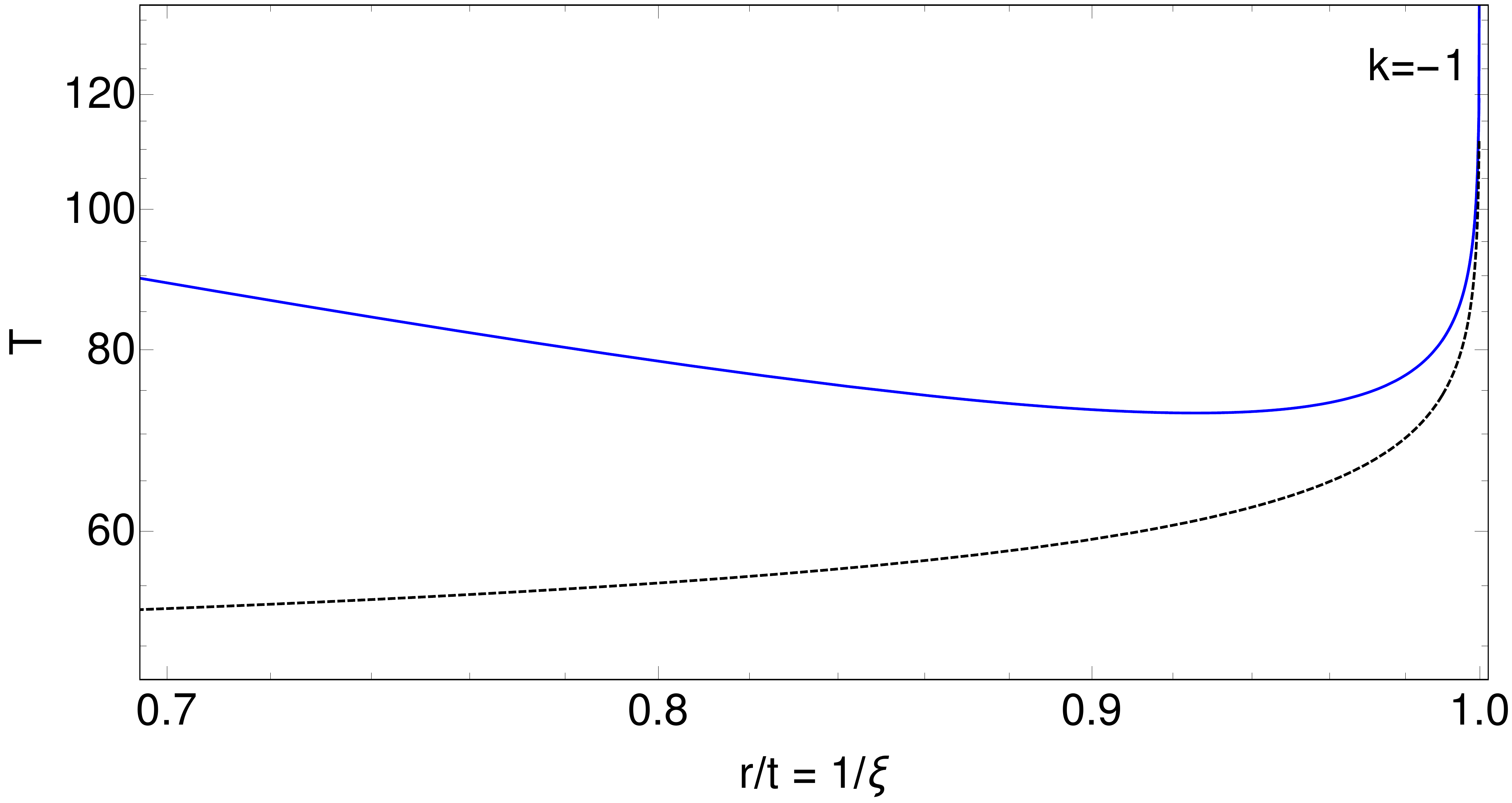} 
\includegraphics[width=1\columnwidth]{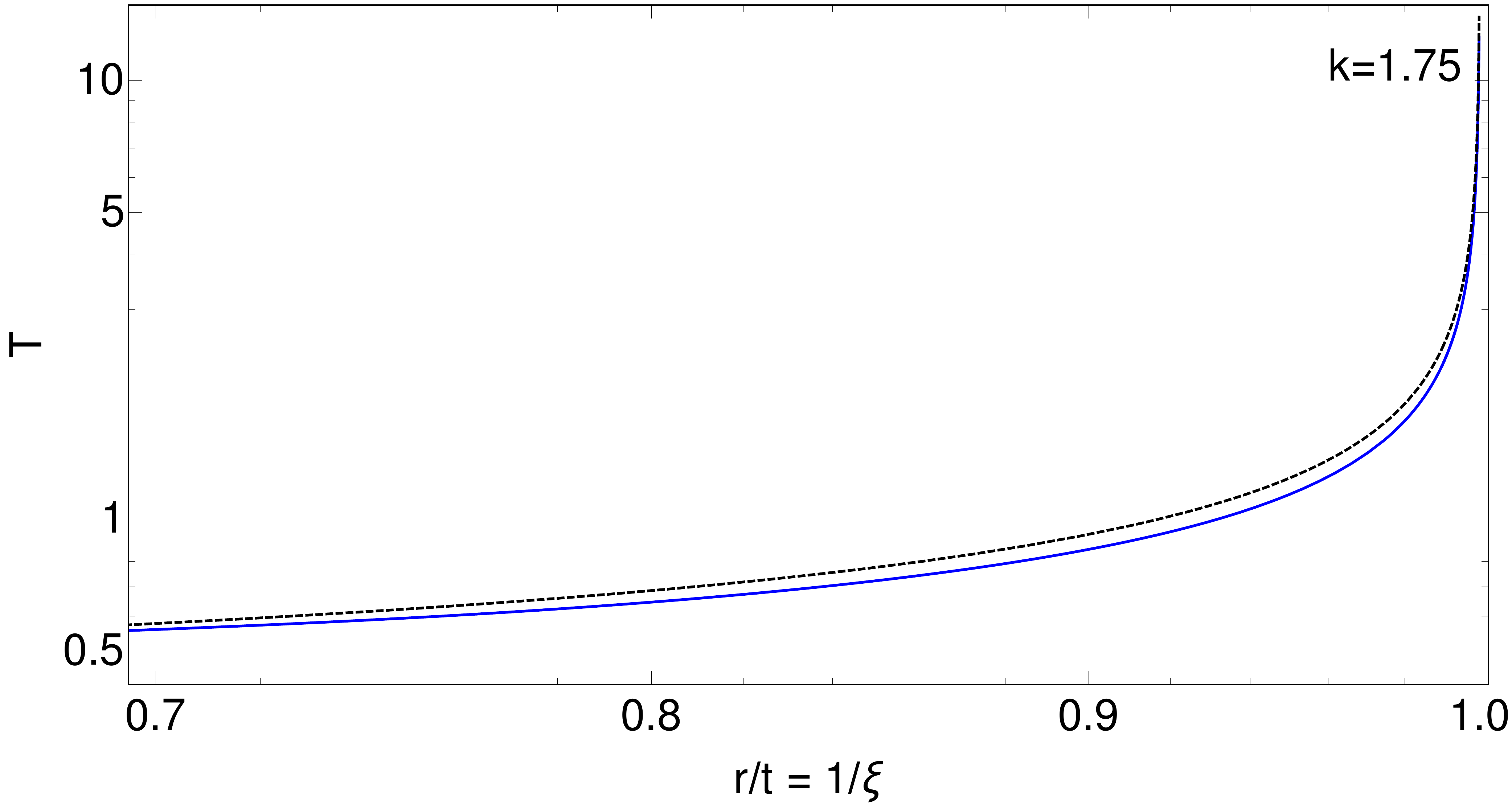} 
\caption{The temperature profiles for $k=-1$ (upper panel) and $k=1.75$ (lower panel) with $\Gamma=10^3$. The Newtonian extension and BM solutions are plotted as solid blue and dashed black lines, respectively. While the temperature profile in the BM solutions decreases behind the shock for every $k<4$, in the Newtonian extension the fluid is reheated towards $r=0$ if $k<3/2$.} \label{f:Temperature}
\end{figure}

In Figure \ref{f:Temperature} we show the temperature profile of the fluid for $k=-1$ and $k=1.75$, with a shock Lorentz factor of $\Gamma=10^3$. While the temperature in the BM solution always decreases behind the shock, in the Newtonian extension the flow becomes hotter far from the shock for $k=-1$, but decreases all the way to the origin for $k=1.75$.

In cases where the gas in the internal solution cools behind the shock, the assumption of an ultra-relativistic equation of state needs to be relaxed. This treatment is beyond the scope of this paper, but see Pan \& Sari\cite{Pan09}. Instead, in this work, we consider the solution as valid only down to where $p/\rho\sim 1$.
\tf{Nevertheless, since the inner flow contains very little energy, the cold part of the flow is unlikely to affect the hot part of the solution.}

\section{Discussion} \label{s:discussion}
We study the interiors of an outgoing, ultra-relativistic shock wave in spherical geometry, allowing for a general density profile of the form $\rho \propto r^{-k}$, where $k<4$, in accordance with the requirement for energy conserving, first-type self-similar solutions in spherical geometry.
We show that our self-similar solutions are successful in describing the flow behind an outgoing shock wave at a distance $\Delta r \gg R/\Gamma^2$ behind the shock, both when the flow is ultra-relativistic, all the way to the sub-relativistic regime at $r \ll R$. It is straightforward to show from the Blandford-McKee solution for $\gamma$ that at a distance of $\Delta r \gg R/\Gamma^2$ behind the shock, the Lorentz factor of the flow no longer depends on the shock Lorentz factor. Thus, for every value of $k$ there exists a unique solution for $\beta$ that does not depend on the shock Lorentz factor $\Gamma$.

Despite being an extension to first type self-similar solutions, the new solutions include a singular point, a feature normally associated with second type solutions. However, in contrast to the latter, the new internal solution does not necessarily pass smoothly through the singular point. For $k<\frac{1}{2}(5-\sqrt{10})\cong 0.92$, the solution is discontinuous, which implies the appearance of a secondary shock wave, while for $\frac{1}{2}(5-\sqrt{10})<k<4$ the solution smoothly crosses the sonic line and no secondary shock appears. According to the temporal dependence of the pressure, we show that the fluid at $r\rightarrow 0$ can either be in-flowing or out-flowing, where the condition for an inflow at the center coincides with the formation of the outgoing secondary shock wave. 
We have shown in section \ref{s:solutions_at_r0} that for $k<\frac{1}{2}(5-\sqrt{10})\cong 0.92$ there is an inflow of material towards the origin. 

Contrary to the BM solutions , in which the temperature behind the shock decreases for every $k<4$, we find that in the interior part of the blast wave the flow can be re-heated towards the origin if $k<3/2$.

\tf{An examination of Figures \ref{f:f_k} and \ref{f:h_k} reveals that in some cases the pressure and density gradients have opposite signs. For example, in the $k=0$ case, the pressure profile increases towards the origin, while the density profile decreases, leading to instability. Close to the origin, the instabilities will grow on the shock dynamical time scale, and will not have enough time to develop significantly. However, in the vicinity of the secondary shock wave, the gradients are larger and the instability will grow on a shorter time scale. In this region, a detailed stability analysis is required.
}

\tf{It is known that if the blast wave is generated by an initial flow of a very high Lorentz factor, the flow will relax to the BM solutions once an energy equivalent to that carried by the shock is given to the medium. However, the interior flow is at much larger scales behind the shock compared to the flow described by the BM solutions, and the timescale for achieving self-similarity is expected to be significantly longer. Self-similarity in the internal flow assumes that the fluid has information about the ultra-relativistic shock wave. 
This requires that a $C_-$ characteristic emerging from $\xi_\sh = 1+1/[2(m+1)\Gamma^2]$ can reach all the way to the origin while the shock is still ultra-relativistic. The time it takes a  $C_-$ characteristic to reach the origin depends on $\Gamma_0$ and is equal to
\begin{equation}
    \frac{t_-}{t_0} \simeq 2.7 ~ \Gamma_0^{0.07} ~
\end{equation}
in the $k=0$ case, while the time the shock becomes non relativistic is $\sim \Gamma_0^{2/3}$. Therefore, a $C_-$ characteristic emerging from the relativistic shock always reaches the origin when the shock is still relativistic and the internal self-similar solution exists.
 We illustrate this in Figure \ref{f:Cplus_char_shock}.}

\tf{Both the BM solution and the extension found in this work assume spherical symmetry. Nevertheless, the BM solution was applied to non-spherical systems, such as ultra-relativistic jets in the outflow of GRBs. The application of the BM solutions in these cases is justified because relativistic beaming of the outflow makes the flow behave like an angular patch of a spherical blast wave. However, the same argument cannot be applied to the Newtonian interior of the blast wave, which means that this solution can be used only in strictly spherical configurations.}

\tf{Our analysis of the interior of the solutions, allows us to discuss questions like the inflow of material on a possible compact remnant located at the origin. Consider a compact object of mass $M$ at $r=0$ that accretes the material around the origin.
For a subsonic flow, accretion effectively occurs through the Bondi radius. Taking an ultra-relativistic speed of sound $c_\s = c/\sqrt{3}$, the Bondi radius is comparable to the Schwarzschild radius of the object. The mass accretion rate in this case is
\begin{equation}
    \dot{m} = 4\pi r^2_\b c_\s \rho_\b \propto t^{\frac{45-12k}{2 k^2-17k+36}}~,
\end{equation}
where $r_\b$ is the Bondi radius and $\rho_\b$ is the density at that radius. Assuming a spherical shock wave propagating into a circumstellar medium with a uniform ambient density $\rho$ (where $k=0$), then
\begin{equation} \begin{split}
    \dot{m} \sim & \frac{\big(GM\big)^{3} \rho^{7/4} c^{15/4}}{E^{3/4} c_\s^5}~ t^{5/4} \sim 10^{-39}  \frac{M_\odot}{\text{yr}}\times \bigg(\frac{M}{M_\odot}\bigg)^{3} \\ & \times \bigg(\frac{E}{10^{52}\text{erg}}\bigg)^{-3/4}\bigg(\frac{n_0}{\text{cm}^{-3}}\bigg)^{7/4}\bigg(\frac{t}{\text{yr}}\bigg)^{5/4},
    \end{split}
\end{equation}
where $\n_0$ is the ambient number density and $E$ is the total energy deposited in the medium by the shock. The accretion rate becomes smaller at larger energies since the fluid's density decreases with increasing $\Gamma$ at the transition to Newtonian velocities. For typical physical parameters, the central object practically does not increase its mass.
}

\section{Acknowledgements}
We thank  Almog Yalinewich and Nicholas Stone for insightful discussions. \tf{We thank the anonymous referee for helpful comments on the manuscript. } This work was supported by an ISF grant.

\section{Data Availability Statement}
Data sharing is not applicable to this article as no new data were created or analyzed in this study.

\section{References}
% Create the reference section using BibTeX:
\bibliography{BM_continuation.bib}

\end{document}